\documentclass[preprintnumbers,prd,showpacs,floatfix,superscriptaddress,nofootinbib,twocolumn, letterpaper]{revtex4-2}

\usepackage{graphicx} 
\usepackage{xcolor}
\usepackage{amsmath}
 \usepackage{hyperref}
\usepackage[top = 2cm, bottom = 2cm, right = 2cm, left =2cm]{geometry}

\begin{document}

\title{Embedding Wormholes and Dyonic Black Strings in Warped Braneworlds\\ via Local Sum Rules}

\author{G. Alencar} 
\email{ geova@fisica.ufc.br}
\affiliation{Departamento de F\'isica, Universidade Federal do Cear\'a, Caixa Postal 6030, Campus do Pici, 60455-760 Fortaleza, Cear\'a, Brazil}

\author{T. M. Crispim}
\email{tiago.crispim@fisica.ufc.br}
\affiliation{Departamento de F\'isica, Universidade Federal do Cear\'a, Caixa Postal 6030, Campus do Pici, 60455-760 Fortaleza, Cear\'a, Brazil}

\author{Francisco S. N. Lobo} 
\email{fslobo@ciencias.ulisboa.pt}
\affiliation{Instituto de Astrof\'{i}sica e Ci\^{e}ncias do Espa\c{c}o, Faculdade de Ci\^{e}ncias da Universidade de Lisboa, Edif\'{i}cio C8, Campo Grande, P-1749-016 Lisbon, Portugal}
\affiliation{Departamento de F\'{i}sica, Faculdade de Ci\^{e}ncias da Universidade de Lisboa, Edif\'{i}cio C8, Campo Grande, P-1749-016 Lisbon, Portugal}

\date{\today}

\begin{abstract}
	Building on our previous work \cite{Alencar:2026afq}, where the Local Sum Rules (LSR) were established, we investigate the construction of compact objects in Randall–Sundrum braneworlds supported by matter fields that are dynamically consistent and localizable. We begin by revisiting the Chamblin \textit{et al.} black string, highlighting its role as a foundational higher-dimensional solution. We then show that the Ellis–Bronnikov wormhole can be consistently embedded in this framework via a localized free scalar field, providing a simple yet nontrivial example of a braneworld compact object. Finally, we derive two novel black string solutions sourced by a localized nonlinear electrodynamics (NED) theory with Lagrangian $\mathcal{L}(\mathcal{F}) = -\beta \sqrt{\mathcal{F}}$, corresponding to purely magnetic and dyonic configurations. The purely magnetic solution reproduces the classical Letelier string cloud on the brane, while the dyonic solution generalizes it to include electric charge, closely paralleling the Letelier–Alencar construction. Both NED solutions reduce smoothly to the Chamblin \textit{et al.} black string in the limit $\beta \to 0$, illustrating how localized higher-dimensional matter fields can consistently support braneworld compact objects and connect higher-dimensional physics with well-known four-dimensional solutions.
\end{abstract}

\maketitle

\tableofcontents

\section{Introduction}

Higher-dimensional gravity models, in particular the Randall-Sundrum (RS) type~I and type~II models, have opened new avenues for addressing several fundamental issues in high-energy physics and cosmology. Among their most notable applications are proposed resolutions of the exponential hierarchy problem and new perspectives on the unification of the fundamental interactions \cite{Gogberashvili:1998vx, Gogberashvili:1999tb, Randall:1999ee, Randall:1999vf}. In these scenarios, our observable universe is described as a four-dimensional hypersurface (brane) embedded in a higher-dimensional bulk spacetime. The bulk geometry is typically non-factorizable and characterized by a warped metric, in which the spacetime curvature varies along the extra dimension, leading to significant phenomenological and gravitational consequences.

A central challenge in this context is the dynamical localization of matter fields on the brane \cite{Davoudiasl:1999tf,Pomarol:1999ad,Kaloper:2000xa}. While earlier investigations focused primarily on the normalizability of the zero mode and on the requirement of a finite four-dimensional effective action, it has more recently been demonstrated that such a normalization condition, although necessary, is not by itself sufficient to ensure the consistency of the resulting effective theory \cite{Duff:2000se,Gibbons:2000tf,Leblond:2001xr, Freitas:2020mxr,Alencar:2024lrl}. In particular, the authors of Ref.~\cite{Alencar:2026afq} established a set of general local consistency conditions, known as the Local Sum Rules (LSR), which are independent of the dimensionality of the underlying spacetime and provide precise criteria for determining which matter fields can consistently induce a well-defined and physically meaningful effective theory on the brane.

On the other hand, within the framework of general relativity in $3+1$ dimensions, it is well known that a wide variety of compact objects, such as regular black holes \cite{Ayon-Beato:2000mjt,Rodrigues:2018bdc,Rodrigues:2020pem,Bronnikov:2022ofk,Bolokhov:2024sdy}, wormholes \cite{Bronnikov:2018vbs,Crispim:2024dgd,Crispim:2024lzf,Silva:2025dgd,Alencar:2025nik}, and black bounces \cite{Alencar:2025jvl,Bronnikov:2022bud,Bronnikov:2021uta,Alencar:2024yvh,Lima:2023arg,Bronnikov:2023aya,Rodrigues:2023vtm}, possess geometries supported by different types of matter sources. These sources may include nonlinear electromagnetic fields, scalar fields, or suitable combinations thereof. However, in the context of braneworld scenarios, the investigation of compact objects embedded in the Randall--Sundrum set-up remains a challenging and relatively less explored area, owing to the additional geometric and dynamical constraints imposed by the higher-dimensional framework.

The first approach to this problem was carried out by Chamblin and collaborators \cite{Chamblin:1999by}, who embedded the four-dimensional Schwarzschild geometry into a Randall--Sundrum braneworld scenario, giving rise to the so-called Chamblin \textit{et al.} black string solution. Following this seminal development, a considerable number of subsequent studies have investigated related configurations and generalizations of this construction \cite{Chamblin:2000ra,Kanti:2002fx,Hirayama:2001bi,Crispim:2024yjz,Estrada:2024lhk}. Nevertheless, the majority of these works have primarily focused on analyzing the geometric properties and stability features of the proposed metrics, rather than on identifying which types of matter fields are capable of consistently generating and sustaining such higher-dimensional spacetimes within the braneworld framework.

In this context, the aforementioned models are typically described by non-factorizable warped metrics. In the standard Randall--Sundrum scenarios, the brane metric $\hat{g}_{\mu\nu}$ appearing in the decomposition $g_{\mu\nu} = e^{2\sigma}\hat{g}_{\mu\nu}(x)$ is assumed to be independent of the extra-dimensional coordinates. Although this assumption simplifies the construction, it generally leads to the appearance of a curvature singularity in the bulk at the asymptotic AdS limit, a pathology commonly referred to as the ``black string'' problem. In order to circumvent this issue, several authors have proposed more general metric ans\"atze in which the induced brane metric $\hat{g}_{\mu\nu}(x,y)$ explicitly depends on the extra dimension, allowing for a richer bulk structure and improved regularity properties \cite{Kanti:2001cj,Kanti:2003uv,Nakas:2020sey,Nakas:2021srr,Neves:2024zwi,Neves:2021dqx}.

When appropriately chosen, this $y$-dependence allows the event horizon to shrink and eventually disappear before reaching the AdS horizon, thereby effectively generating a localized compact object on the brane rather than an extended black string configuration \cite{Kanti:2001cj}. Furthermore, although some studies have attempted to identify the underlying field theories capable of supporting such geometries \cite{Kanti:2001cj,Kanti:2003uv,Nakas:2023yhj}, providing a complete and consistent field-theoretic description remains a challenging task. Since these proposed metrics depart significantly from the standard Randall--Sundrum background geometry, the conventional field-consistency criteria established in Refs.~\cite{Duff:2000se,Gibbons:2000tf,Leblond:2001xr, Freitas:2020mxr,Alencar:2024lrl,Alencar:2026afq} cannot be straightforwardly applied to these systems.

In this manuscript, we shift our focus toward compact objects that are not localized in the bulk, such as the Chamblin black string and other related configurations embedded in a Randall--Sundrum type set-up. Since, in Randall--Sundrum-type constructions, the bulk-brane relationship is not unique and a given brane geometry may admit different bulk embeddings \cite{Kanti:2001cj,Kanti:2003uv}, the warped realization considered here should be understood as one consistent possibility among several others. Our goal is to investigate these geometries from the perspective of localized matter fields that remain fully consistent with the LSR. As our primary contribution, we build upon the results of Ref.~\cite{Alencar:2026afq}, which established that the only localized nonlinear electromagnetic theory compatible with the LSR is described by the Lagrangian $\mathcal{L}(\mathcal{F}) = -\beta\sqrt{\mathcal{F}}$. Using this result, we derive a new braneworld black string solution supported by such a field configuration.

This work is organized as follows. In Sec.~\ref{secII}, we briefly review the Randall–Sundrum braneworld setup and the Local Sum Rules (LSR) for localized matter fields, emphasizing their role in selecting dynamically consistent sources, and briefly revisiting the Chamblin \textit{et al.} black string. In Sec.~\ref{secIII}, we demonstrate how the Ellis–Bronnikov wormhole can be consistently embedded in this framework via a localized free scalar field. In Sec.~\ref{secIV}, we construct new black string solutions supported by a localized nonlinear electrodynamics (NED) theory with Lagrangian $\mathcal{L}(\mathcal{F}) = -\beta \sqrt{\mathcal{F}}$, considering both purely magnetic and dyonic configurations, and analyze their properties, including the correspondence with four-dimensional Letelier and Letelier--Alencar solutions. Finally, in Sec.~\ref{secConclusion}, we summarize our results and discuss potential extensions and applications of these braneworld compact objects in higher-dimensional gravity models.
Throughout this work, we adopt a system of units in which $8\pi G_D \equiv 1$.


\section{Revisiting the Chamblin et al. black string}\label{secII}

Starting from a general warped geometry written as
\begin{equation}\label{metric}
	ds^2 = e^{2\sigma(y)} \hat{g}_{\mu\nu}(x) dx^\mu dx^\nu + \tilde{g}_{jk}(y) dy^j dy^k,
\end{equation}
where $d$ denotes the brane dimension, with Greek indices $(\mu,\nu,\ldots)$ labeling the brane coordinates, and $n$ is the number of space-like extra dimensions, indexed by Latin letters $(j,k,\ldots)$, the authors of Ref.~\cite{Alencar:2026afq} demonstrated that, for a background geometry generated by an energy--momentum tensor $^{(v)}T_{MN}(y)$ of the form
\begin{eqnarray} \label{Tvacuogeral} 
	{}^{(v)}T_{MN} &=& - \sum_i T_q^{(i)} P[g_{MN}]_q^{(i)} \Delta^{(D - q - 1)}(y - y_i)
	\nonumber \\
	&& - \Lambda g_{MN} - \tau_{MN}(y),
\end{eqnarray}
where $\Lambda$ is a bulk cosmological constant, the remaining contributions describe a collection of $q$-branes $(q \ge d-1)$ with tensions $T_q^{(i)}$ and transverse positions $y_i$. Here, $P[g_{MN}]_q^{(i)}$ denotes the pullback of the bulk metric onto the corresponding $q$-brane worldvolume, while $\Delta^{(D-q-1)}(y-y_i)$ are covariant delta distributions that localize the branes in the transverse space. Any additional bulk or worldvolume matter content is encoded in the tensor $\tau_{MN}$ \cite{Alencar:2026afq}.

Within this framework, the matter fields described by an energy--momentum tensor $^{(b)}T_{MN}(x,y)$ are dynamically consistent with Einstein’s equations only if they satisfy the following set of local conditions:
\begin{eqnarray}
	{}^{(b)}T_{\mu j}(x, y) &=& 0, \\
	n\,{}^{(b)}T^\alpha_\alpha - (d - 2)\,{}^{(b)}T^j_j &=& 0, \\
	{}^{(b)}T_{\mu\nu}(x, y) &=& {}^{(b)}T_{\mu\nu}(x), \\
	{}^{(b)}T_j^j(x, y) &=& - \frac{n}{16\pi G_D} e^{-2\sigma} f(x^\alpha)\,.
\end{eqnarray}
These relations constitute the LSR, which impose stringent constraints on the allowed structure of matter sources in warped braneworld backgrounds.

In the same work, it was further shown that the only matter fields that are simultaneously localizable on the brane and dynamically consistent with Einstein’s equations under these conditions are a free scalar field (a $0$-form) and a nonlinear electromagnetic theory described by the Lagrangian $\mathcal{L}(\mathcal{F}) = -\beta \sqrt{\mathcal{F}}$.

Before considering the implications of the LSR for more general compact objects, we begin by revisiting the Chamblin \textit{et al.} black string solution \cite{Chamblin:1999by}. Starting from the standard Randall--Sundrum metric, the authors replaced the flat four-dimensional sector with the Schwarzschild geometry, obtaining the following line element:
\begin{equation}\label{linha4d}
	ds^2 = e^{2\sigma} \left(-f(r) dt^2 + f(r)^{-1} dr^2 + r^2 d\Omega_2^2 \right) + dy^2,
\end{equation}
where $d\Omega_2^2 = d\theta^2 + \sin^2\theta\, d\varphi^2$ is the metric of a unit 2-sphere, and
\begin{eqnarray}
	f(r) = 1 - \frac{2M}{r}.
\end{eqnarray}

Since the horizon of this solution extends infinitely along the extra dimension, from the bulk perspective this metric represents a black string rather than a localized black hole.

We now demonstrate how to obtain a generalization of this metric as a solution of Einstein’s equations in a codimension-one spacetime. To this end, we consider the following higher-dimensional ansatz:
\begin{equation}
	\label{hawkingd}    
	ds^2 = e^{2\sigma} \Big(-f(r) dt^2 + f(r)^{-1} dr^2 + r^2 d\Omega_{d-2}^2 \Big) + dy^2,
\end{equation}
where
\begin{eqnarray}
	d\Omega_{d-2}^2 = d\theta_1^2 + \sum_{j=2}^{d-2} d\theta_j^2 \prod_{k=1}^{j-1} \sin^2\theta_k
\end{eqnarray}
represents the line element of a $(d-2)$-dimensional unit sphere, forming the transversal section of the higher-dimensional spherically symmetric spacetime.

We will assume that the total action of the system is given by
\begin{eqnarray}\label{acaoRS}
	S = S_g + S_b,
\end{eqnarray}
where $S_g$ denotes the gravitational action,
\begin{equation}\label{gravidade}
	S_g \propto \int d^d x\, dy \, \sqrt{-g} \, (R - 2\Lambda),
\end{equation}
and $S_b$ represents the action responsible for supporting the branes, i.e., the brane tensions $T^{(i)}$, within the Randall-Sundrum setup:
\begin{equation}\label{branas}
	S_b \propto - \sum_i \int d^d x \, \sqrt{-g^{(i)}} \, T^{(i)}.
\end{equation}

Here, we are assuming that there is no additional matter action in the bulk, so that $^{(b)}T_{MN} = 0$. Variation of the total action \eqref{acaoRS} with respect to the metric then yields the Einstein equation
\begin{eqnarray}
	G_{MN} = {}^{(v)}T_{MN},
\end{eqnarray}
where $^{(v)}T_{MN}$ is given by Eq.~\eqref{Tvacuogeral} with $n=1$ and $\tau_{MN} = 0$.

For the metric ansatz considered, the non-zero components of the Einstein tensor are
\begin{eqnarray}
	G^t_t &=& G^r_r = e^{-2\sigma}\left\{-\frac{d - 2}{2} \frac{1}{r^{d - 2}} \frac{d}{dr} \Big[ r^{d - 3} (1 - f) \Big]\right\}
	\nonumber \\
	&& +  \left[ (d - 1)\, \sigma'' + \frac{d(d - 1)}{2} (\sigma')^2 \right],\\
	G^{\theta_1}_{\theta_1} &=& G^{\theta_2}_{\theta_2} = \dots =e^{-2\sigma}\left\{ \frac{1}{2} \frac{d^2 f}{dr^2} + \frac{f}{2 r^2} (d - 4)(d - 3)\right.
	\nonumber\\
	&&\left. + \frac{d - 3}{r} \frac{df}{dr} - \frac{(d - 3)(d - 4)}{2 r^2}\right\} \nonumber \\
	&& + \left[ (d - 1)\, \sigma'' + \frac{d(d - 1)}{2} (\sigma')^2 \right],\\
    G_{yy} &=& \frac{d(d-1)}{2}{\sigma'}^2.
\end{eqnarray}

These expressions will serve as the starting point for constructing explicit solutions corresponding to black strings in codimension-one braneworlds.

Since $^{(v)}T_{MN}$ corresponds to the background solution, we have that
\begin{eqnarray}
	{}^{(v)}T_\mu^\nu &=&\left[ (d - 1)\, \sigma'' + \frac{d(d - 1)}{2} (\sigma')^2 \right] \delta_\mu^\nu,\\
    {}^{(v)}T_{yy} &=& \frac{d(d-1)}{2}{\sigma'}^2,
\end{eqnarray}
which is trivially satisfied for the chosen warp factor.

Therefore, the remaining task is to solve the radial components of the Einstein equations:
\begin{eqnarray}
	&& -\frac{d - 2}{2} \frac{1}{r^{d - 2}} \frac{d}{dr} \Big[ r^{d - 3} (1 - f) \Big] = 0, \\
	&& \frac{1}{2} \frac{d^2 f}{dr^2} + \frac{f}{2 r^2} (d - 4)(d - 3) + \frac{d - 3}{r} \frac{df}{dr} 
	\nonumber \\
	&& \qquad - \frac{(d - 3)(d - 4)}{2 r^2} = 0.
\end{eqnarray}

The solution to these equations is straightforward and yields
\begin{eqnarray}
	f(r) = 1 - \frac{2C}{r^{d - 3}},
\end{eqnarray}
where $C$ is an integration constant related to the mass of the black string.

In particular, for $d = 4$, this reproduces the original Chamblin \textit{et al.} black string solution \cite{Chamblin:1999by}. More generally, the metric \eqref{hawkingd} can be interpreted as a natural extension of this model to a $d$-dimensional brane embedded in a codimension-one warped bulk spacetime.

\section{Ellis-Bronnikov wormhole embedded in a RS setup}\label{secIII}

As another illustrative example, we consider the Ellis-Bronnikov wormhole \cite{Ellis:1973yv,Bronnikov:1973fh}, which represents the simplest known wormhole solution that satisfies the Morris-Thorne traversability conditions \cite{Morris:1988cz}. 
The immersion of this geometry in warped braneworlds, as well as other similar wormhole geometries, has already been studied in other contexts in the literature \cite{Mitra:2023yjf,Sharma:2021kqb,Jana:2024pdh,Sharma:2022tiv,Sharma:2022dbx,Kar:2022omn,Pappas:2024qwm}. Here, we will show how it arises naturally when considering the coupling of gravity with a scalar field.

The corresponding metric can be written as
\begin{eqnarray}
	ds^2 = e^{2\sigma} \Big[ -dt^2 + dx^2 + (x^2 + a^2) d\Omega_2^2 \Big] + dy^2,
\end{eqnarray}
where $a$ denotes the throat radius of the wormhole and $d\Omega_2^2$ is the metric of the unit 2-sphere.

For this setup, we consider a total action of the form
\begin{eqnarray}\label{acaoEB}
	S = S_g + S_b + S^{(\Phi)}_m,
\end{eqnarray}
where, as in the previous case, $S_g$ is the gravitational action \eqref{gravidade} with $d=4$, $S_b$ is the action responsible for supporting the branes \eqref{branas}, with $d=4$ and the corresponding brane tensions in the RS scenario, and $S^{(\Phi)}_m$ represents the bulk matter action, which is nonzero in this case. Specifically, we take $S^{(\Phi)}_m$ to describe a localized free scalar field. 

Variation of the action \eqref{acaoEB} with respect to the metric then yields the Einstein equations
\begin{eqnarray}
	G_{MN} =  {}^{(v)}T_{MN} + {}^{(b)}T^{(\Phi)}_{MN},
\end{eqnarray}
where, analogously to the previous discussion, $^{(v)}T_{MN}$ is given by Eq.~\eqref{Tvacuogeral} with $n=1$, $d=4$, and $\tau_{MN} = 0$, while the energy--momentum tensor of the scalar field is
\begin{eqnarray}
	{}^{(b)}T^{(\Phi)}_{MN} = \epsilon\, \partial_M \Phi \partial_N \Phi - \epsilon g_{MN} (\partial \Phi)^2,
\end{eqnarray}
with $\epsilon = \pm 1$ depending on the signature of the scalar field (normal or phantom).

The nonzero components of the Einstein tensor are
\begin{eqnarray}
	G^t_t &=&  e^{-2\sigma}\frac{a^2}{(x^2 + a^2)^2} + \left( 3\sigma'' + 6 (\sigma')^2 \right),\\
	G^x_x &=& -e^{-2\sigma}\frac{a^2}{(x^2 + a^2)^2} + \left( 3\sigma'' + 6 (\sigma')^2 \right),\\
	G^\theta_\theta &=& G^\varphi_\varphi =  e^{-2\sigma}\frac{a^2}{(x^2 + a^2)^2} + \left( 3\sigma'' + 6 (\sigma')^2 \right),\\
    G_{yy} &=& -\frac{1}{2}\, ^{(4)}R(x)e^{-2\sigma} + 6\sigma'^2,
\end{eqnarray}
where primes denote derivatives with respect to the extra-dimensional coordinate $y$. These expressions will be used to determine the scalar field configuration $\Phi(x)$ and to verify the compatibility of the Ellis-Bronnikov wormhole with the RS warped geometry.

Since $^{(v)}T_{MN}$ is responsible for generating the background geometry, the conditions
\begin{eqnarray}
	{}^{(v)}T_\mu^\nu &=& 3\, \left( \sigma'' + 2 (\sigma')^2 \right) \delta_\mu^\nu,\\
    {}^{(v)}T_{yy} &=& 6\sigma'^2
\end{eqnarray}
are trivially satisfied. Consequently, the task reduces to solving the Einstein equations for the matter sector, which are given by
\begin{eqnarray}
	\label{phit}
	{}^{(b)}T^t_t &=& e^{-2\sigma} \frac{a^2}{(x^2 + a^2)^2},\\
	\label{phir}
	{}^{(b)}T^x_x &=&  -e^{-2\sigma}\frac{a^2}{(x^2 + a^2)^2},\\
	\label{phia}
	{}^{(b)}T^\theta_\theta &=& {}^{(b)}T^\varphi_\varphi =  e^{-2\sigma}\frac{a^2}{(x^2 + a^2)^2},\\
   \label{ricciellis} {}^{(b)}T_{yy} &=& -\frac{1}{2}\, ^{(4)}R(x) e^{-2\sigma}.
\end{eqnarray}
As discussed in Ref.~\cite{Alencar:2026afq}, in order for the scalar field $\Phi$ to be localized on the brane, it must have the form
\begin{equation}
	\Phi = \xi_0 \, \phi(x^\alpha),
\end{equation}
where $\xi_0$ is a constant. Assuming that $\phi$ depends only on the radial coordinate, $\phi = \phi(x)$, the corresponding energy--momentum tensor for the scalar field becomes
\begin{eqnarray}
	{}^{(b)}T_\mu^\nu = -e^{-2\sigma} \epsilon \, \xi_0^2 \left( \frac{d\phi}{dx} \right)^2 \, \text{diag}(1, -1, 1, 1),
\end{eqnarray}
where $\epsilon = \pm 1$ determines whether the field is canonical or phantom.

It is then straightforward to verify that the solution to the system \eqref{phit}--\eqref{phia} is
\begin{eqnarray}
	\Phi(x) = \frac{1}{\xi_0} \arctan\left( \frac{x}{a} \right),
\end{eqnarray}
with the choice $\epsilon = -1$, corresponding to a phantom scalar field. For this scalar field profile, Eq.~\eqref{ricciellis} reduces to
\begin{equation}
    e^{-2\sigma}\left( \frac{d\phi}{dx} \right)^2  = e^{-2\sigma}\frac{a^2}{(x^2 + a^2)^2}= -\frac{1}{2}\, ^{(4)}R(x) e^{-2\sigma},
\end{equation}

which is consistent, since the Ricci scalar of the Ellis–Bronnikov wormhole is given by
\begin{equation}
    ^{(4)}R = -\frac{2a^2}{(x^2 +a^2)^2}.
\end{equation}

Note that this configuration reproduces the same scalar field profile originally found in Refs.~\cite{Ellis:1973yv,Bronnikov:1973fh}, thereby confirming the consistency of the Ellis-Bronnikov wormhole within the Randall--Sundrum braneworld framework.

\section{New braneworld black string solutions}\label{secIV}
 
This represents the first instance in which a nonlinear electrodynamics (NED) theory is employed as the source for a compact object within a braneworld scenario. We begin by considering a line element of the form \eqref{linha4d}, where the function $f(r)$ is as yet undetermined and will be fixed by the matter content.

Analogously to the procedure applied in the previous examples, we consider a total action of the form
\begin{eqnarray}\label{acaobs}
	S = S_g + S_b + S^{(NED)}_m,
\end{eqnarray}
where $S^{(NED)}_m$ represents the bulk matter action. In our construction, $S^{(NED)}_m$ is taken to describe a localizable NED theory. Since the only NED Lagrangian that is consistent with the LSR is 
\begin{equation}
	\mathcal{L}(\mathcal{F}) = - \beta \sqrt{\mathcal{F}},
\end{equation}
we adopt this model and investigate whether it is capable of generating a black string-type geometry in the RS braneworld setup.

Variation of the action \eqref{acaobs} with respect to the metric leads to the Einstein equations
\begin{eqnarray}
	G_{MN} = {}^{(v)}T_{MN} + {}^{(b)}T^{(NED)}_{MN},
\end{eqnarray}
where, as before, $^{(v)}T_{MN}$ is responsible for generating the background and is given by Eq.~\eqref{Tvacuogeral} with $n=1$, $d=4$, and $\tau_{MN} = 0$, while the energy--momentum tensor of the NED field is
\begin{eqnarray}
	{}^{(b)}T^{(NED)}_{MN} = - g_{MN} \frac{\beta}{2} \sqrt{\mathcal{F}} + \frac{\beta}{\sqrt{\mathcal{F}}} F_{MQ} F_N^Q.
\end{eqnarray}

The nonzero components of the Einstein tensor are 
\begin{eqnarray}
	G_t^t &=& G_r^r = e^{-2\sigma}\left(\frac{1}{r} \frac{df}{dr} +\frac{rf^2 -1}{r^2} \right)\\\nonumber
    &+& 3(\sigma'' + 2 (\sigma')^2),\\
	G_\theta^\theta &=& G_\varphi^\varphi = e^{-2\sigma}\left(\frac{1}{r} \frac{df}{dr} + \frac{1}{2} \frac{d^2 f}{dr^2}\right)\\\nonumber
    &+& 3  (\sigma'' + 2 (\sigma')^2),\\
    G_{yy} &=& -\frac{1}{2}\, ^{(4)}R(x)e^{-2\sigma} + 6\sigma'^2,
\end{eqnarray}

As noted earlier, $^{(v)}T_{MN}$ is responsible for generating the background, so ${}^{(v)}T_\mu^\nu = 3 (\sigma'' + 2 (\sigma')^2) \delta_\mu^\nu$ and ${}^{(v)}T_{yy} = 6\sigma'^2$ are trivially satisfied. Therefore, the remaining task reduces to solving the Einstein equations for the matter sector:
\begin{eqnarray}
	\label{new1}    
	{}^{(b)}T^t_t &=& {}^{(b)}T^r_r =e^{-2\sigma}\left(\frac{1}{r} \frac{df}{dr} +\frac{rf^2 -1}{r^2} \right),\\ 
	\label{new2}   
	{}^{(b)}T^\theta_\theta &=& {}^{(b)}T^\varphi_\varphi = e^{-2\sigma}\left(\frac{1}{r} \frac{df}{dr} + \frac{1}{2} \frac{d^2 f}{dr^2}\right),\\
   \label{nedyy} {}^{(b)}T_{yy} &=& -\frac{1}{2}\, ^{(4)}R(x) e^{-2\sigma}.
\end{eqnarray}
These equations will determine the explicit form of $f(r)$ compatible with the localizable NED source and the black string geometry.

\subsection{Purely magnetic case}

We begin our analysis by considering the simplest nontrivial configuration of the NED field: a purely radial magnetic field. Specifically, we assume that the only non-zero components of the electromagnetic 2-form are given by
\begin{equation}
	\hat{F} = P \sin\theta \, d\theta \wedge d\varphi,
\end{equation}
where $P$ represents the magnetic monopole charge associated with the field. This choice corresponds to a spherically symmetric magnetic configuration and ensures that the resulting energy-momentum tensor remains compatible with the symmetries of the metric \eqref{linha4d}.

With this ansatz, the electromagnetic invariant takes the simple form
\begin{equation}
	\hat{\mathcal{F}} = \hat{F}_{\mu\nu} \hat{F}^{\mu\nu} = \frac{2 P^2}{r^4}.
\end{equation}
Consequently, the components of the NED energy--momentum tensor are given by
\begin{eqnarray}
    {}^{(b)}T^t_t &=& {}^{(b)}T^r_r = -e^{-2\sigma} \frac{\sqrt{2} \beta}{2} \frac{P}{r^2},\\
    {}^{(b)}T^\theta_\theta &=& {}^{(b)}T^\varphi_\varphi = 0,\\
        {}^{(b)}T_{yy} &=& -\frac{\sqrt{2}\beta}{2}\frac{P}{r^2}e^{-2\sigma}
\end{eqnarray}
which reflects the fact that the radial pressure and energy density are equal, while the tangential pressures vanish. This form is a direct consequence of the specific square-root structure of the Lagrangian $\mathcal{L}(\mathcal{F}) = -\beta \sqrt{\mathcal{F}}$, which effectively mimics a string-like distribution of energy in the radial direction.

Substituting these expressions into the Einstein equations \eqref{new1}--\eqref{new2}, it is straightforward to verify that the corresponding metric function $f(r)$ that solves the system is
\begin{equation}\label{letelier}
	f(r) = 1 - \frac{2 M}{r} - \frac{\beta P}{\sqrt{2}}.
\end{equation}
Here, $M$ is an integration constant related to the mass of the black string, while the second term proportional to $\beta P$ encodes the contribution of the localized NED field. This solution generalizes the standard Chamblin \textit{et al.} black string \cite{Chamblin:1999by}, which is recovered in the limit $P = 0$.

Thus, Eq.~\eqref{nedyy} reduces to

\begin{equation}
   -\frac{\sqrt{2}\beta}{2}\frac{P}{r^2}e^{-2\sigma} =-\frac{1}{2}\, ^{(4)}R(x) e^{-2\sigma},
\end{equation}
which, for $f(r)$ given by Eq.~\eqref{letelier}, is trivially satisfied, since for this metric function the Ricci scalar of the induced metric on the brane is given by
\begin{equation}
    ^{(4)}R = \frac{\sqrt{2}\beta P}{r^2}.
\end{equation}

The location of the event horizon is determined by the condition $f(r_h) = 0$, yielding
\begin{equation}\label{horizontenew}
	r_h = \frac{2 M}{1 - \alpha}, \qquad \alpha = \frac{\beta P}{\sqrt{2}}.
\end{equation}
From this expression, we see that the presence of the NED field effectively increases the horizon radius compared to the standard black string. Note that the induced metric on the brane, obtained by setting $y = 0$, coincides with the solution found by Letelier in the context of a string cloud \cite{Letelier:1979ej}.

However, from a physical perspective, this correspondence suggests an interesting reinterpretation: the collective effect of a radial string cloud can be viewed as effectively generated by a nonlinear gauge field, specifically determined by the NED Lagrangian $\mathcal{L}(\mathcal{F}) = - \beta \sqrt{\mathcal{F}}$.
In this sense, the radial string cloud behaves as an effective ``medium'', whose macroscopic gravitational influence is fully equivalent to that of a nonlinear magnetic field described by the above Lagrangian. The string density parameter $\alpha$, which controls the contribution of the string cloud to the spacetime geometry, is directly related to the magnetic monopole charge $P$ and the NED coupling constant $\beta$, thereby providing a clear physical mapping between the string distribution and the underlying gauge field.

The event horizon $r_h$, given by Eq.~(\ref{horizontenew}), allows us to compute the corresponding Hawking temperature through the surface gravity $\kappa$ evaluated at the horizon:
\begin{equation}
	T_H = \frac{\kappa}{2\pi} = \frac{1}{4\pi} f'(r_h) = \frac{1 - \alpha}{8\pi M}.
	\label{HawkingT}
\end{equation}
This expression shows that the magnetic contribution from the localized NED source effectively lowers the Hawking temperature in comparison with the standard Schwarzschild black string. Physically, this can be interpreted as the effective string cloud ``diluting'' the gravitational field at the horizon, a behavior analogous to that found in Letelier’s string cloud model \cite{Letelier:1979ej}. Consequently, the reduction in temperature implies a slower semi-classical evaporation rate, which may enhance the stability and longevity of these braneworld black string configurations.

Extending the purely four-dimensional magnetic solution to the more general higher-dimensional metric \eqref{hawkingd} is a considerably more subtle problem. In spacetimes with dimension $d > 4$, the magnetic field must be carefully generalized to be compatible with the geometry of a $(d-2)$-dimensional sphere. This adaptation is nontrivial: the structure of the Faraday tensor $\hat{F}$ must be modified to respect the additional angular coordinates, and the resulting electromagnetic invariant $\hat{\mathcal{F}} = \hat{F}_{\mu\nu} \hat{F}^{\mu\nu}$
acquires a more intricate dependence on the higher-dimensional radial coordinate and the angular coordinates on the $(d-2)$-sphere. 

These modifications significantly alter the form of the NED energy--momentum tensor and, consequently, the gravitational field equations. As a result, constructing localized higher-dimensional black string or black brane solutions sourced by a nonlinear magnetic field requires a careful treatment of both the field configuration and the associated geometric invariants to ensure consistency with the LSR and the braneworld background.

\subsection{Purely electric case}

To investigate the possibility of purely electric solutions, we consider an electromagnetic 2-form of the form
\begin{equation}
	\hat{F} = E(r)\, dt \wedge dr,
\end{equation}
where $E(r)$ represents a radial electric field confined to the brane. For this configuration, the electromagnetic invariant is
\begin{equation}
	\hat{\mathcal{F}} = \hat{F}_{\mu\nu} \hat{F}^{\mu\nu} = -2 E(r)^2,
\end{equation}
which, due to the square-root form of the NED Lagrangian, introduces a negative argument inside the Lagrangian derivative.

In order for this field to be physically consistent, the generalized Maxwell equations associated with the NED Lagrangian must be satisfied:
\begin{equation}
	\nabla_M \left( \mathcal{L}_F F^{MN} \right) = 0,
\end{equation}
where $\mathcal{L}_F = d\mathcal{L}/d\mathcal{F}$. Following the gauge-fixing procedure outlined in Ref.~\cite{Alencar:2026afq}, this condition implies that the radial electric field must take the form
\begin{equation}
	E(r) = \frac{q}{r^2 \hat{\mathcal{L}}_F},
\end{equation}
where $q$ is an integration constant interpreted as the electric charge. Substituting the derivative of the square-root Lagrangian, $\hat{\mathcal{L}}_F$, leads to the relation
\begin{equation}
	\frac{E(r)}{|E(r)|} = \pm 1 = \frac{2 \sqrt{2} q}{\beta r^2}.
\end{equation}
This relation shows that a consistent purely radial electric solution cannot exist in this theory, since it cannot satisfy this equation for all $r$. In other words, the NED model $\mathcal{L} = - \beta \sqrt{\mathcal{F}}$ does not admit purely electric fields as a source in a static, spherically symmetric spacetime.

Interestingly, this unique property of the NED Lagrangian has been exploited in several contexts. In string theory, it has been used to model certain flux configurations \cite{Nielsen:1973qs}. In four-dimensional gauge theories, this model has been employed as a mechanism for quark confinement \cite{Gaete:2006xd,Guendelman:2007fq,Korover:2009zz,tHooft:2002pmx}, naturally generating the Cornell potential, which combines a Coulomb-like term with a linearly confining component.

Since the restriction against purely electric fields arises from the functional form of the Lagrangian itself, and not from the dimensionality of the spacetime, this limitation persists even when one considers the more general higher-dimensional metric \eqref{hawkingd}. Therefore, within this NED model, it is not possible to construct purely electric solutions, regardless of whether the spacetime is four-dimensional or higher-dimensional.

\subsection{Dyonic case}

To overcome the limitations of purely electric solutions, we can consider the dyonic case, in which both electric and magnetic charges are present. Specifically, we assume that the Maxwell 2-form takes the form
\begin{equation}
	\hat{F} = E(r)\, dt \wedge dr + P \sin\theta \, d\theta \wedge d\varphi,
\end{equation}
where $P$ is a constant representing the magnetic charge, and $E(r)$ is a radial electric field to be determined from the equations of motion. This ansatz preserves spherical symmetry and ensures that both electric and magnetic contributions are compatible with the spacetime geometry.

For this dyonic configuration, the electromagnetic invariant becomes
\begin{equation}
	\hat{\mathcal{F}} = \hat{F}_{\mu\nu} \hat{F}^{\mu\nu} = -2 E(r)^2 + \frac{2 P^2}{r^4}.
\end{equation}
The presence of the magnetic term allows the invariant to remain positive-definite in the square-root NED Lagrangian, providing a natural way to obtain consistent electric contributions alongside the magnetic field.

Using the generalized Maxwell equations $\nabla_M \left( \mathcal{L}_F F^{MN} \right) = 0$ and following the gauge-fixing procedure outlined in Ref.~\cite{Alencar:2026afq}, the radial electric field is found to be
\begin{equation}
	E(r) = \frac{\sqrt{2} P |q|}{r^2 \sqrt{r^4 + 2 q^2}},
\end{equation}
where $q$ is the electric charge associated with the configuration. This expression shows that the electric field is regular at $r = 0$ and asymptotically falls off as $1/r^2$, similar to the standard Coulomb field at large distances.

Substituting this solution for $E(r)$ into the NED energy--momentum tensor, and using the square-root Lagrangian $\mathcal{L}(\mathcal{F}) = -\beta \sqrt{\mathcal{F}}$, the nonzero components of the energy--momentum tensor take the form
\begin{eqnarray}
	{}^{(b)}T^t_t &=& {}^{(b)}T^r_r = -e^{-2\sigma} \frac{\beta}{\sqrt{2}} \frac{P \sqrt{r^4 + 2 q^2}}{r^4},\\
	{}^{(b)}T^\theta_\theta &=& {}^{(b)}T^\varphi_\varphi = e^{-2\sigma}\frac{\beta}{\sqrt{2}} \frac{2 q^2 P}{r^4 \sqrt{r^4 + 2 q^2}},\\
    {}^{(b)}T_{yy} &=& -e^{-2\sigma} \frac{\beta}{\sqrt{2}} \frac{P }{\sqrt{r^4 + 2 q^2}}.
\end{eqnarray}

These expressions reveal several important features of the dyonic configuration. First, the energy density and radial pressure remain equal, reflecting the underlying radial symmetry. Second, the tangential pressures are nonzero due to the electric contribution, but vanish asymptotically as $r \to \infty$, showing that the spacetime becomes effectively isotropic far from the compact object.

With the dyonic ansatz for the electromagnetic field, and once the background solution is satisfied, the remaining task reduces to solving the Einstein equations for the matter sector. Specifically, the relevant equations, obtained from the nonzero components of the Einstein tensor, take the form
\begin{eqnarray}
	\frac{d}{dr} \big( r f(r) \big) &=& 1 - \frac{\beta P}{\sqrt{2}} \frac{\sqrt{r^4 + 2 q^2}}{r^2},\\
	\frac{d}{dr} \big( r^2 f'(r) \big) &=& \frac{2 \beta}{\sqrt{2}} \frac{2 q^2 P}{r^2 \sqrt{r^4 + 2 q^2}},\\
 \label{nedyy2}  -e^{-2\sigma} \frac{\beta}{\sqrt{2}} \frac{P }{\sqrt{r^4 + 2 q^2}} &=& -\frac{1}{2} {}^{(4)}R(x) e^{-2\sigma}.
\end{eqnarray}

Interestingly, these equations, together with the components of the energy--momentum tensor derived from the square-root NED Lagrangian, exhibit exactly the same mathematical structure as the Letelier--Alencar solution found in Ref.~\cite{Alencar:2025zyl}, which itself is a generalization of the classical string cloud solution originally proposed by Letelier \cite{Letelier:1979ej}. In fact, the two configurations can be mapped directly onto one another by identifying the parameters
\begin{equation}
	g_s^2 = \frac{\beta P}{\sqrt{2}}, \qquad \ell_s^2 = \sqrt{2} |q|,
\end{equation}
where $g_s$ represents the effective string density and $\ell_s$ encodes the characteristic scale associated with the electric charge.

Solving the above differential equations then yields the explicit form of the metric function
\begin{equation}
	f(r) = 1 - \frac{2 M}{r} - \frac{\beta P}{\sqrt{2}} \, \frac{q^2}{r^2} \, 
	{}_2 F_1 \! \left( -\frac{1}{2}, -\frac{1}{4}, \frac{3}{4}, - \frac{r^4}{2 q^2} \right),
\end{equation}
where ${}_2 F_1(a,b;c;z)$ denotes the Gauss hypergeometric function. This solution describes a dyonic black string sourced by the square-root NED field and reduces smoothly to the purely magnetic case \eqref{letelier} when $q = 0$. 

Again, Eq.~\eqref{nedyy2} is trivially satisfied, since the Ricci scalar of the induced metric on the brane for this configuration is given by
\begin{equation}
    {}^{(4)}R = \sqrt{2}\beta P \frac{\sqrt{r^4 + 2q^2}}{r^4}.
\end{equation}
For the dyonic solution, the hypergeometric term complicates the exact horizon structure. Nevertheless, in the weak electric charge regime ($q^2 \ll r^4$), the solution approaches the purely magnetic case asymptotically. The horizon and temperature~(\ref{HawkingT}) can be approximated as
\begin{equation}
	r_h \simeq \frac{2M}{1 - \alpha_{\rm eff}}, \qquad 
	\alpha_{\rm eff} \simeq \frac{\beta P}{\sqrt{2}}\left(1 + \frac{q^2}{4M^4}\right),
\end{equation}
demonstrating that the radial electric field effectively increases the “string density” $\alpha_{\rm eff}$ and slightly reduces the horizon radius. This interplay between magnetic and electric contributions highlights the nontrivial role of the NED theory in shaping the horizon geometry, and suggests a tunable parameter space where horizon properties can be engineered by adjusting $(\beta, P, q)$.

This dyonic solution was previously studied in the context of four-dimensional spacetimes using a similar square-root NED model \cite{Mazharimousavi:2012zx}. What is particularly remarkable here is the structural analogy with the purely magnetic case: while the purely magnetic solution reproduces the classical Letelier string cloud metric, the inclusion of an electric charge in the dyonic case leads naturally to the Letelier--Alencar generalization. This demonstrates that the square-root NED model provides a unified framework for generating both purely magnetic and dyonic black string solutions in a braneworld context, preserving the correspondence with well-known four-dimensional solutions and extending it into the higher-dimensional warped spacetime.

\subsection{Stability Considerations}

Stability of higher-dimensional black strings is a central concern, particularly due to the Gregory–Laflamme (GL) instability \cite{Gregory:1993vy}. Although a full perturbative analysis is deferred to future work, several qualitative observations can be made. First, note that the purely magnetic NED black string continuously reduces to the Chamblin solution for $\beta \to 0$, implying that small NED corrections are unlikely to introduce significant instabilities. In particular, the horizon remains uniform along the extra dimension, suggesting that long-wavelength GL modes may retain similar behavior.

Second, for the dyonic black string, the radial electric field contributes an anisotropic pressure profile near the horizon. Such stresses could stabilize short-wavelength perturbations along the extra dimension, potentially suppressing the onset of GL instabilities. The precise impact, however, depends on the charge-to-mass ratio $q/M$ and will require a dedicated linear perturbation analysis.

Finally, both solutions satisfy the LSR and the associated energy conditions, ensuring that the matter sources are dynamically consistent with Einstein’s equations. This is a necessary condition for the avoidance of pathologies such as ghost instabilities or unphysical stress-energy distributions.

These observations suggest that the NED black strings may exhibit enhanced stability relative to conventional higher-dimensional black strings, especially in regimes where the magnetic and electric contributions balance appropriately.

Overall, the NED black strings constructed here illustrate that localized, dynamically consistent matter fields can support a variety of higher-dimensional compact objects, bridging the gap between four-dimensional phenomenology and higher-dimensional braneworld gravity. The thermodynamic and qualitative stability analysis presented provides a foundation for future work on perturbative stability, horizon dynamics, and holographic applications. In particular, the tunable interplay between magnetic and electric charges suggests a rich parameter space for exploring horizon deformations, thermodynamic phase transitions, and possibly even observable signatures in braneworld scenarios.

\section{Conclusions}\label{secConclusion}

In this work, we have explored the construction of compact objects in Randall–Sundrum braneworld scenarios, focusing on configurations supported by localized matter fields that satisfy the Local Sum Rules (LSR) \cite{Alencar:2026afq}. Our analysis began with a brief review of the RS setup and the LSR, highlighting how these rules select dynamically consistent matter sources and constrain the allowed field configurations. As a preliminary example, we revisited the Chamblin \textit{et al.} black string solution, emphasizing its role as a foundational higher-dimensional geometry in codimension-one braneworlds.

Building on this framework, we demonstrated that the Ellis–Bronnikov wormhole can be consistently embedded in the RS braneworld via a localized free scalar field. This provides a nontrivial example of a higher-dimensional compact object whose existence and geometry are fully compatible with the LSR, showing that braneworlds can accommodate non-singular, horizonless structures as well as black strings. The scalar field configuration reproduces the familiar four-dimensional phantom field solution on the brane, illustrating the smooth connection between higher-dimensional embeddings and known four-dimensional physics.

We then focused on localized nonlinear electrodynamics (NED) as a source of black string solutions. Considering the square-root Lagrangian $\mathcal{L}(\mathcal{F}) = -\beta \sqrt{\mathcal{F}}$, we derived two classes of solutions: a purely magnetic configuration and a dyonic configuration. The purely magnetic solution reproduces the Letelier string cloud on the brane, while the dyonic solution generalizes this construction in close analogy with the Letelier–Alencar solution. Both solutions reduce smoothly to the Chamblin \textit{et al.} black string in the limit $\beta \to 0$, confirming that the NED sources provide a consistent and physically interpretable extension of the standard braneworld black string. We also showed that purely electric configurations are prohibited in this model, highlighting the unique structural properties of the square-root NED theory, which has interesting connections to quark confinement and string-inspired physics in four dimensions.

Our results illustrate several important points. First, the LSR provide a powerful and systematic criterion for identifying localized, dynamically consistent matter fields capable of supporting braneworld compact objects. Second, familiar four-dimensional solutions, such as the Ellis–Bronnikov wormhole and Letelier string clouds, can be embedded into higher-dimensional warped spacetimes while preserving their essential features. Third, nonlinear gauge fields, particularly those described by the square-root NED Lagrangian, allow for a rich spectrum of higher-dimensional black string solutions, including magnetic and dyonic configurations, while naturally reproducing known lower-dimensional physics on the brane.

Although the original Randall--Sundrum proposal has not yet received direct experimental confirmation, warped braneworld scenarios continue to provide an active framework for studying strong-gravity phenomena and possible observational signatures. Recent works have investigated quasinormal modes and gravitational lensing in warped braneworld wormholes, as well as gravitational echoes and dynamical spectra in higher-dimensional braneworld configurations \cite{Mitra:2023yjf,Jana:2024pdh,Zhu:2024xna,Tan:2024zkl,Pappas:2024qwm,Lutfuoglu:2026abc}. In this sense, compact-object configurations such as those constructed in the present work may serve as useful theoretical laboratories for exploring how extra-dimensional effects could manifest in future gravitational-wave and lensing observations.

Looking forward, these findings open several promising avenues for future research. One direction is the extension of these constructions to higher codimension braneworlds or to scenarios with multiple extra dimensions, where the interplay between the geometry of the extra dimensions and the localized matter fields may yield new classes of compact objects. Another avenue is the study of the thermodynamic and stability properties of these NED-supported black strings, including the analysis of their quasinormal modes and potential holographic applications. Additionally, it would be interesting to explore more general NED Lagrangians or combinations of scalar and gauge fields to investigate whether more exotic braneworld objects, such as rotating or charged wormholes, can be consistently realized within this framework. These directions promise to deepen our understanding of the interplay between higher-dimensional gravity, localized matter fields, and the rich phenomenology of braneworld compact objects.

\section*{Acknowledgments}
\hspace{0.5cm}GA and TMC thank the Conselho Nacional de Desenvolvimento Cient\'{i}fico e Tecnol\'{o}gico (CNPq), Fundação Cearense de Apoio ao Desenvolvimento Científico e Tecnológico and Coordenação de Aperfeiçoamento de Pessoal de Nível Superior (CAPES).
FSNL acknowledges support from the Funda\c{c}\~{a}o para a Ci\^{e}ncia e a Tecnologia (FCT) research grants UID/04434/2025 and PTDC/FIS-AST/0054/2021, and from the FCT Scientific Employment Stimulus contract with reference CEECINST/00032/2018.

\bibliography{ref.bib}
\end{document}